\documentclass{article}

\usepackage{PRIMEarxiv}

\usepackage[utf8]{inputenc} 
\usepackage[T1]{fontenc}    
\usepackage{hyperref}       
\usepackage{url}            
\usepackage{booktabs}       
\usepackage{amsfonts}       
\usepackage{nicefrac}       
\usepackage{microtype}      
\usepackage{lipsum}
\usepackage{fancyhdr}       
\usepackage{graphicx}       
\graphicspath{{media/}}     

\usepackage{amssymb}
\usepackage{pifont}
\newcommand{\cmark}{\ding{51}}%
\newcommand{\xmark}{\ding{55}}%
\usepackage{subcaption}

\title{Multi-input architecture and disentangled representation learning for multi-dimensional modeling of music similarity
\thanks{Submitted to ICASSP 2022}
}

\author{
  Sebastian Ribecky, Jakob Abeßer, Hanna Lukashevich \\
  Semantic Music Technologies Group \\
  Fraunhofer IDMT \\
  Ilmenau, Germany \\
  \texttt{\{sebastian.ribecky, jakob.abesser, hanna.lukashevich\}@idmt.fraunhofer.de} \\
}

\begin{document}
\maketitle

\begin{abstract}
In the context of music information retrieval, similarity-based approaches are useful for a variety of tasks that benefit from a query-by-example scenario. Music however, naturally decomposes into a set of semantically meaningful factors of variation. Current representation learning strategies pursue the disentanglement of such factors from deep representations, resulting in highly interpretable models. This allows the modeling of music similarity perception, which is highly subjective and multi-dimensional. While the focus of prior work is on metadata driven notions of similarity, we suggest to directly model  the human notion of multi-dimensional music similarity. To achieve this, we propose a multi-input deep neural network architecture, which simultaneously processes mel-spectrogram, CENS-chromagram and tempogram in order to extract informative features for the different disentangled musical dimensions: genre, mood, instrument, era, tempo, and key. We evaluated the proposed music similarity approach using a triplet prediction task and found that the proposed multi-input architecture outperforms a state of the art method. Furthermore, we present a novel multi-dimensional analysis in order to evaluate the influence of each disentangled dimension on the perception of music similarity.
\end{abstract}


\section{Introduction}

Music information retrieval (MIR) approaches traditionally aim to automatically annotate music recordings with textual annotations that describe their musical attributes. Such annotations form \textit{semantic spaces} within data collections, which structure the music library and allow its efficient browsing.

However, the retrieval of songs based on these semantic spaces requires a relatively high musical understanding by part of the user in its query. In most scenarios, arguably, the user is not really sure how to textually describe the desired musical traits, but has an example in another song of the similar characteristics or ``feel'' searched for. In such cases of higher subjectivity, a query-by-example approach is convenient, as it simplifies the retrieval task for the user. Another application scenario is music replacement, where a seed musical track needs to be replaced during the production process, for instance due to the licensing restrictions, but still a similar musical track is wanted. For this purposes, \textit{music similarity-based retrieval} algorithms are developed.

Moreover, music similarity is a highly subjective concept, which is considered inherently \textit{multi-dimensional} \cite{RN2}. Different musical pieces can be considered as similar (or dissimilar) over one or more dimensions of the different facets they present, such as their specific music characteristics (e.g. rhythm, tempo, key, melody, instrumentation), semantic conceptualizations (e.g. genre, style, theme), perceived moods (e.g. happy, aggressive, calm), listening situations (e.g. to work-out, to study, to party), or musicological factors (e.g. composer, influences), among others. One of the greatest challenges for music similarity algorithms is to identify which combination of these dimensions are most relevant to a particular user when rating two pieces as similar.

\begin{figure}[htbp]
    \centerline{\includegraphics[width=0.6\linewidth]{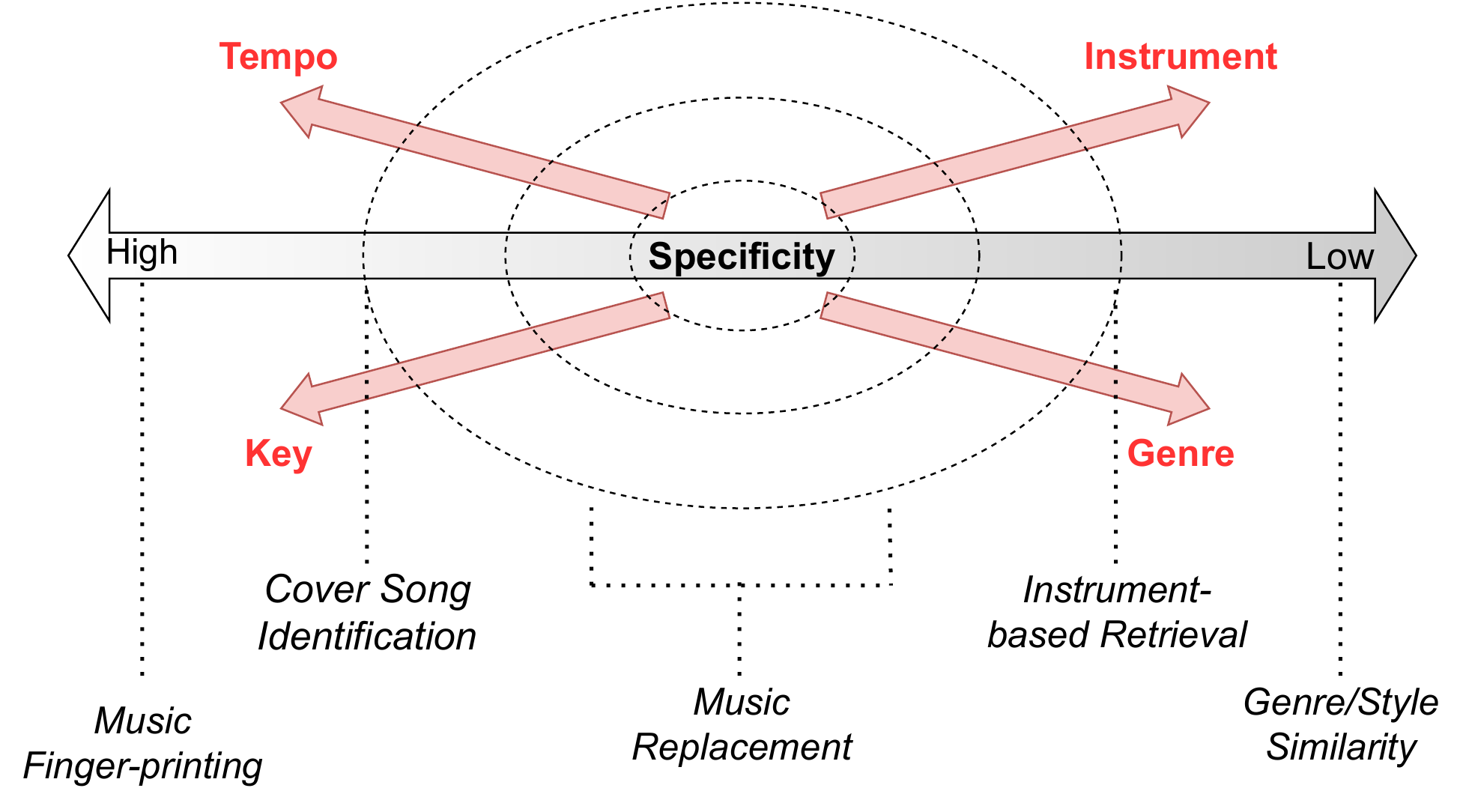}}
    \caption{Specificity spectrum for music similarity-based retrieval tasks (inspired by \cite{RN40}). Examples are allocated along the horizontal axis according to the degree of specificity they convey. The elliptical shapes illustrate the similarity spaces generated by the configurable musical dimensions.}
    \label{fig:spec}
\end{figure}

MIR approaches are currently based on learning efficient hierarchical representations from input data by \textit{deep neural networks}. Recent \textit{representation learning} approaches for music similarity \cite{RN1, RN2} focus on the \textit{disentanglement} of the multi-dimensionality inherent in deep representations. This step will allow to perform a wide range of music retrieval applications with a higher efficiency \cite{RN1} and to configure the extent to which the different disentangled musical dimensions are considered in the latent space of the network. Thus, a single framework can potentially be used to generate adjustable \textit{similarity spaces}, targeting tasks with different degrees of specificity, as shown in Fig.~\ref{fig:spec}. Furthermore, disentangled representations allow the modeling of inter-dimensional relations in user queries. This is of particular relevance for mid/low specificity tasks, in which the notion of similarity is increasingly subjective.

In this paper, we expand on recent work on disentanglement representation learning  \cite{RN1} and employ a triplet-based metric learning approach to disentangle six musical concepts: genre, mood, instrumentation, era, tempo and key. We introduce a novel multi-input deep architecture in order to disentangle these concepts. In this way, we incorporate both low-level and mid-level musical features in the representation space of the network, instead of relying solely on high-level conceptualizations. We employ a perceptual similarity approach, in which triplets containing human annotations of music similarity are used to evaluate different configurations of our proposed model. Additionally, we conduct a novel multi-dimensional analysis to study the influence of the disentangled musical concepts on the human perception of music similarity.

\section{Related Work}

\subsection{Music Similarity \& Deep Metric Learning}

Many tasks in MIR such as music replacement, music recommendation, and music playlist generation fall naturally into a query-by-example setting, where a user queries the system by providing a song or a fragment of it and the system responds by retrieving a list of relevant or similar songs \cite{RN5}. Such applications do not require textual annotations, but ultimately depend on the notion of music similarity between songs to produce high-quality results.

\textit{Deep metric learning} is a representation learning approach, which allows to quantify such abstract notions of similarity. Through a \textit{distance metric}, it aims to map data to an embedding space, in which similar data-points are close together and dissimilar data-points are further away from each other. One common deep metric learning paradigm includes triplets of samples as input in the form of an anchor, a positive (similar to the anchor), and a negative (dissimilar to the anchor) sample. This relationship between samples can be exploited to train a deep architecture \cite{RN10} using an embedding loss, namely the \textit{triplet margin loss} \cite{RN9}. 

Although training may be slower compared to other metric learning paradigms \cite{RN6}, the triplet-based learning approach offers a higher flexibility, since its form of supervision comprises a relative comparison. Recently, sampling strategies to generate triplets that lead to a faster convergence have been proposed, such as semi-hard negative mining \cite{RN7} or distance weighted sampling \cite{RN8}.

\subsection{Disentangled Representation Learning}

Apart from deep metric learning, the idea of learning disentangled representations has also attracted a lot of attention in the machine learning research community in recent years. In general, it is considered that the disentanglement of representations is an important step towards an improved representation learning \cite{RN16, RN17}. In the music domain, disentanglement learning has mostly been developed in the field of music generation for tasks, such as genre transfer \cite{RN19}, timbre and pitch synthesis \cite{RN20}\cite{RN21}, instrument rearrangement \cite{RN22}, musical attributes manipulation \cite{RN23}\cite{RN24}, and rhythm transfer \cite{RN25}\cite{RN26}. Recently however, a disentangled metric learning approach for music similarity has been proposed by Lee et\,al. \cite{RN2}. The authors adapt the \textit{conditional similarity network} (CSN) \cite{RN3} model from the computer vision domain (image similarity) to the music domain. It essentially consists in the expansion of the basic triplet metric learning network \cite{RN10} to include \textit{dimensional masks}, which activate certain portions of the embeddings generated by a \textit{backbone architecture} during training. In this way, equally-sized portions of the embedding space of the network are assigned for metric learning of the different concepts to disentangle.

\section{Proposed Method}

We propose to expand the CSN architecture for music similarity to disentangle six musical dimensions: genre, mood, instrument, era, tempo, and key.

\subsection{Audio Pre-Processing \& Input Features}

We train the network using three second long audio segments of samples from the Million Song Dataset (MSD) \cite{RN31}. Following previous work \cite{RN32, RN1}, we group the Last.FM tags associated with the MSD files into four dimensions: genre, mood, instrument, and era. We use the Madmom Python library \cite{RN27} to extract key \cite{RN29} and tempo \cite{RN28} information from the dataset files.  

Using the librosa library \cite{RN35}, we compute three different input representations of the dataset files: the mel-spectrogram, the cyclic tempogram \cite{RN34}, and the chroma energy normalized statistics (CENS-chromagram) \cite{RN33}. For the mel-spectrograms, a window size of 23~ms with 50~\% overlap is used and 128 mel-bands per frame are computed with logarithmic compression as $log_{10}(1 + 10 * S)$, where $S$ corresponds to the energy-magnitude spectrogram. Additionally, the resulting mel-spectrograms are z-score standardized \cite{RN2}.

For the tempograms and CENS-chromagrams, the same window size and hop size (23~ms and 50~\% overlap, respectively) is used to compute the short-time Fourier transform (STFT). In this way, all input representations have a concurring temporal axis, which simplifies the sampling process. However, the length of the onset auto-correlation window in the tempogram is set to the default value of 384 bins; while the number of chroma for the CENS-chromagram is set to 12 (one for each note of the chromatic scale), computed from a CQT with a resolution of 36 bins per octave.

\subsection{Neural Network Architecture}

\begin{figure}[tbp]
	\centering
	\begin{subfigure}{0.49\linewidth}
		\centering
		\includegraphics[width=0.85\linewidth]{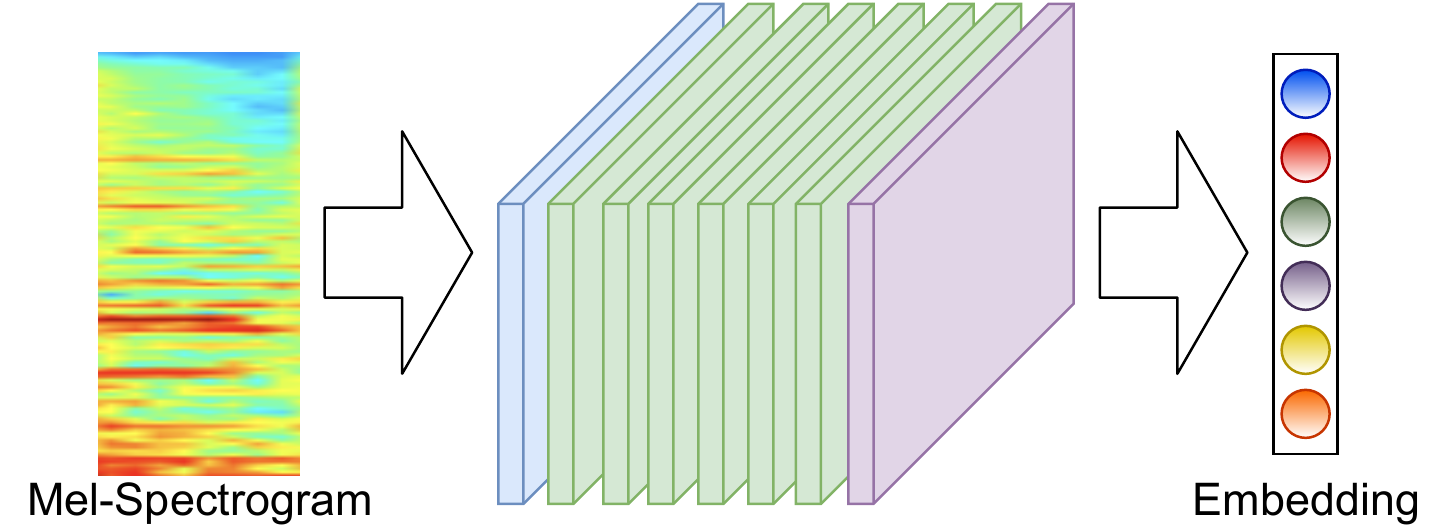}
		\subcaption{Baseline single-input architecture.}
		\label{fig:arch_1}
	\end{subfigure}
	\begin{subfigure}{0.49\linewidth}
		\centering
		\includegraphics[width=0.95\linewidth]{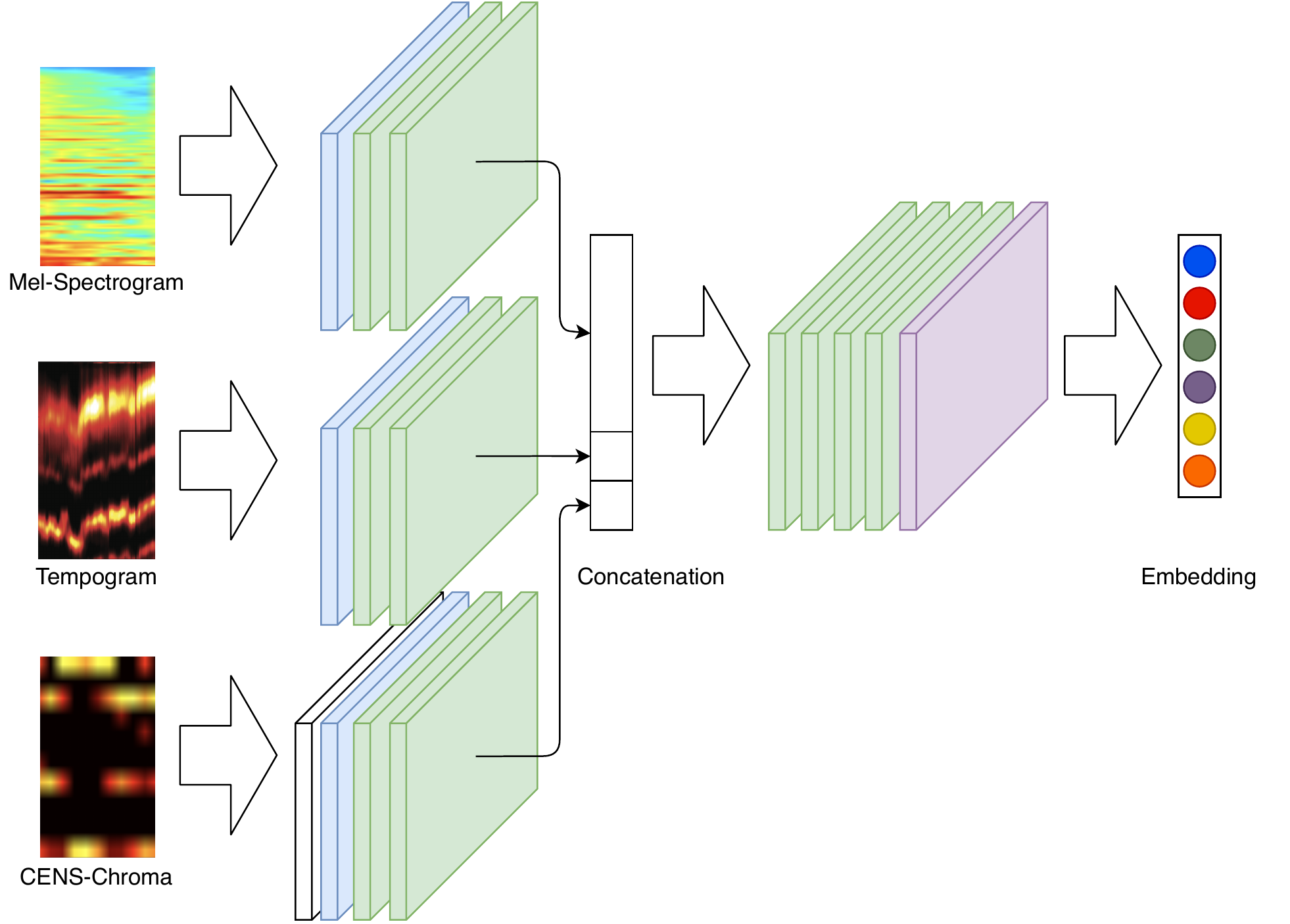}
		\subcaption{Proposed multi-input architecture.}
		\label{fig:arch_2}
	\end{subfigure}
	\caption{Summary of both investigated network architectures. Rectangular blocks denote the different processing units: basic convolutional blocks in blue, inception blocks in green, dense layers in violet, and zero-padding layer in white (used for the CENS-chromagram).} 
	\label{fig:arch}
\end{figure}

Fig.~\ref{fig:arch} provides an overview of the implemented backbone networks. The models are based on the Inception network \cite{RN27} from the computer vision domain. Fig.~\ref{fig:arch_1} shows the baseline backbone network first introduced in \cite{RN2}. This architecture consists of a basic block of convolutional and maxpooling operations processing the single input of the network (mel-spectrogram), followed by six Inception blocks, each consisting of two Inception modules \cite{RN27}: a ``na\"ive'' module, followed by a ``dimension reduction'' module. The final layer is a dense layer with the same number of neurons as the resulting embedding size. Layer normalization \cite{RN36} is applied to the generated embeddings (i.e., embedding-wise normalization) in order to increase efficiency during training \cite{RN39}.

Furthermore, we propose a variant of the baseline embedding architecture (Fig.~\ref{fig:arch_2}). It consists on the parallel processing of the mel-spectrogram, the tempogram and CENS-chromagram of the input samples. Thus, a processing branch is created for each input, conformed by a convolutional layer, and followed by two inception blocks. For the CENS-chromagram branch an additional zero-padding layer is employed to upsample the input along the frequency axis to 16 bins. In this way, the outputs of all branches are 16x16 feature maps, which are concatenated and further processed in the same way as for the baseline model.

\subsection{Conditional Triplet Sampling \& Learning Strategy}

The employed learning method requires sampling triplets of the form 
\begin{equation}
        t_{s} = (x_{a},x_{p},x_{n};s)
    \label{eqn:cs_triplet}
\end{equation}
where $x_{a}$, $x_{p}$, and $x_{n}$ are the anchor, positive and negative samples, respectively, within a considered embedding dimension $s$. We employ a conditional triplet loss $L_s(t_{s})$ \cite{RN2} for each dimension individually, and average them to produce the total multi-dimensional loss of the network $L_\mathrm{MD}$ for $N_\mathrm{D}$ number of dimensions:

\begin{equation}
        L_\mathrm{MD} = \frac{1}{N_\mathrm{D}} \sum_{s=1}^{N_\mathrm{D}} L_s
    \label{eqn:multidim_loss}
\end{equation}

For the triplet sampling, we consider certain notions of similarity. For the dimensions inferred from the dataset labels (genre, mood, instrument, and era), we form anchor-positive pairs from samples that simply present the same label. For tempo, we consider two samples as similar if the extracted global tempo of their respective songs are within a 5 bpm margin (to compensate for possible noise in the tempo estimation algorithm) or if they are any number of tempo-octaves higher or lower, following previous work done on tempo similarity \cite{RN37}. Finally, for key, samples are considered similar if the extracted key of their respective songs are the same, or if they are parallel keys.
Additionally, we employ semi-hard negative mining \cite{RN7} to find negative samples that are close to the anchor, but not as close as the positive sample, within each musical dimension individually. Negatives are mined within a margin value of 0.1 of the anchor-positive distance to avoid easy negatives.

\subsection{Training Parameters}

All model-variations were trained using ``Adam'' optimizer \cite{RN38} with an initial learning rate of $1e^{-5}$, which is reduced by a factor of 5 each time the validation loss does not decrease after 10 epochs, up to a minimum learning rate of $1e^{-10}$. Cosine distance was selected as the distance metric for training.

\section{Evaluation}

\subsection{Dataset \& Evaluation Metric}

In an evaluation phase, we test the performance of the disentangled embeddings obtained from the implemented method (CSNs) for perceptual music similarity. To do so, we employ the recently published Dim-sim dataset \cite{RN2}. It consists of 4,000 triplets sampled from the test split of the MSD, each with human similarity annotations. However, as all similarity annotations are subjective, we make use of a refined split of 426 high-agreement triplets with an agreement rate equal or above 90\%. Embeddings are extracted from three second long non-overlapping sections and averaged over the full song duration in order to obtain the track embeddings used for evaluation.

We use the triplet prediction score as evaluation metric, calculating the fraction of correct predictions given by the evaluated model for the ground-truth triplets in the dataset. Here, the predictions are regarded as correct if the computed euclidean distance in the embedding space between the anchor and positive sample is smaller than that of the anchor and negative. In this way, a measure of concurrence between the trained model predictions of similarity within a triplet, and the human perception of triplet similarity is obtained.

\subsection{Experiment 1 - Model Performance}

To test model performance of different configurations, the entirety of the embedding space is activated (no masks) while processing the evaluation triplets. Tab.~\ref{tab:eval_simil} displays the results obtained. Models with four dimensions include exclusively the disentanglement of the concepts present in the MSD-labels, while six-dimensional models include, additionally, the tempo and key dimensions. The ``baseline'' configuration refers to the CSN model with the standard backbone network and all parameters set mirroring the original study \cite{RN1}. As an additional benchmark, we evaluate the OpenL3-embeddings \cite{RN4}, also averaged over the full song duration in the same manner as the disentangled embedding.

It is observed that solely increasing the number of dimensions with respect to the baseline does not provide an improvement in performance. However, by employing the proposed multi-input architecture and keeping the same number of dimensions, already a slight improvement is observed. The best result is obtained for the six-dimensional multi-input architecture. It further improves the triplet prediction performance compared to all other configurations including the baseline. These observations are indicative that higher-level dimensions (genre, mood, instrument, and era) also make use of the information present in the two additional input representations; and that solely the mel-spectrogram does not provide sufficient informative cues for the tempo and key dimensions.

\begin{table}[tbh!]
    \small
	\centering
	\caption{Results of different model configurations.}
		\begin{tabular}{lcccc}
			\toprule
			\textbf{Model} & \textbf{Number of} & \textbf{Multi-} & \textbf{Embedding} & \textbf{Triplet} \\
			& \textbf{Dimensions} & \textbf{Input} & \textbf{Size} & \textbf{Score} \\
			\midrule
			\textit{Baseline}\cite{RN1} & 4 & \xmark & 256 & 0.8192 \\
			\midrule
            \textit{OpenL3}\cite{RN4} & - & \xmark & 512 & 0.7958 \\
			\midrule
			& 4 & \cmark & 256  & 0.8286 \\
            \textit{Proposed} & 6 & \xmark & 258 & 0.7934 \\
            \textit{Model} & 6 & \xmark & 384 & 0.8169 \\
             & 6 & \cmark & 384 & \textbf{0.8380} \\
			\bottomrule
		\end{tabular}
	\label{tab:eval_simil}
\end{table}

\subsection{Experiment 2 - Multi-Dimensional Analysis of Similarity Perception}

\begin{figure}[tbh!]
    \centerline{\includegraphics[width=0.4\linewidth]{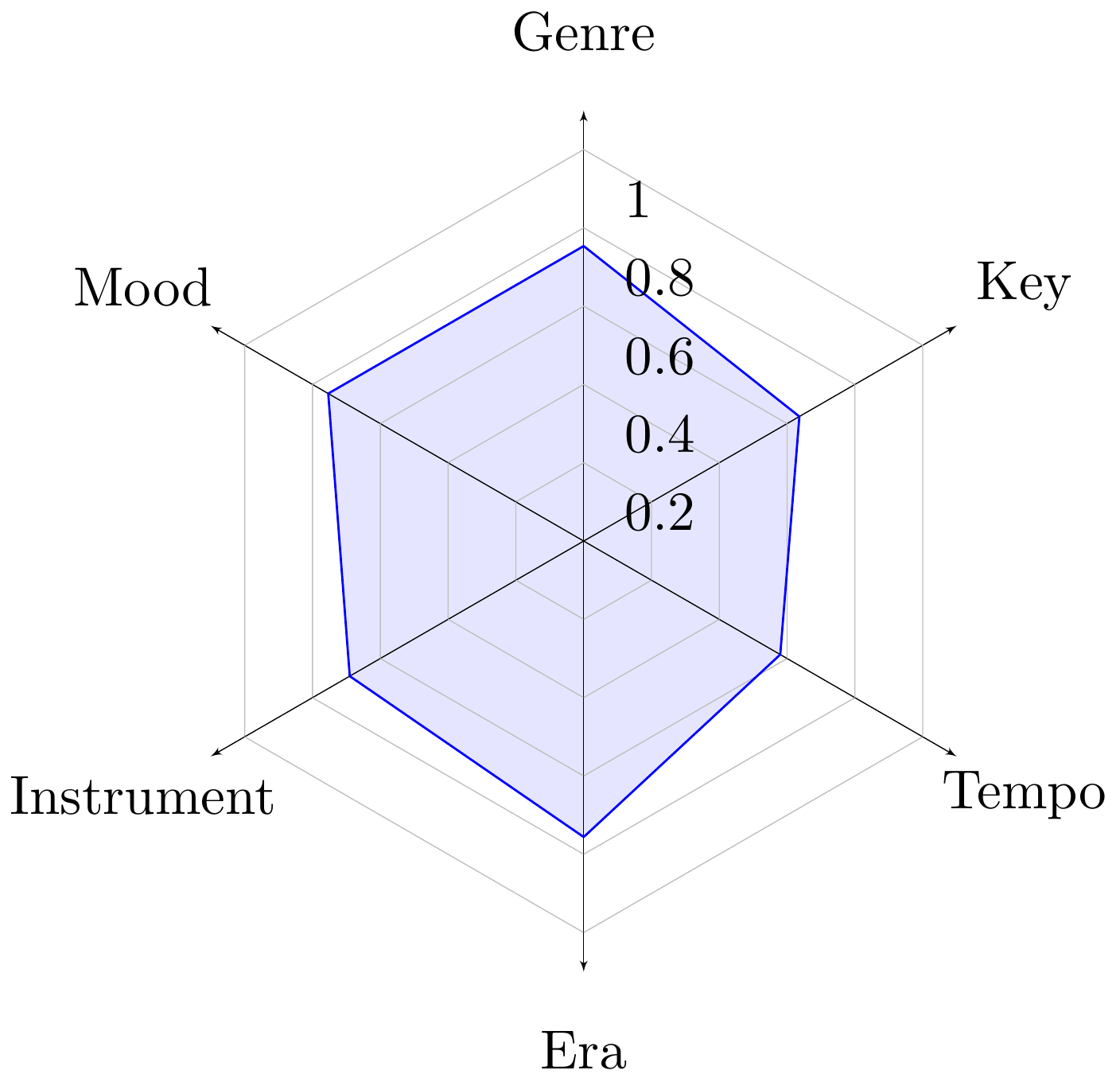}}
    \caption{Influence of each musical dimensions individually.}
    \label{fig:kiviat}
\end{figure}

To carry out the multi-dimensional analysis, the best performing model (see Tab.~\ref{tab:eval_simil}) is evaluated by activating only particular sub-spaces of the produced embeddings, which correspond to the different disentangled concepts. The top half of Tab.~\ref{tab:eval_multidim} shows the top six combinations of disentangled concepts, which yielded the best triplet prediction results; while the bottom half displays the results when activating each individual concept. Figure \ref{fig:kiviat} depicts these last results in a Kiviat diagram, in order to obtain an improved visualization of the influence each individual concept has over the human perception of music similarity, according to the implemented deep model.

\begin{table}[tbh!]
    \small
    \centering
    \caption{Results of the multi-dimensional analysis. A ``\checkmark'' denotes activation of the corresponding dimension.}
	\begin{tabular}{ccccccc}
    	\toprule
    	\multicolumn{6}{c@{}}{\textbf{Dimension}}&\textbf{Triplet}\\
    	\cmidrule(l){1-6}
    	\textbf{Genre}&\textbf{Mood}&\textbf{Instrument}&\textbf{Era}&\textbf{Tempo}&\textbf{Key}&\textbf{Score}\\
    	\midrule
    	\checkmark & \checkmark & \checkmark & \checkmark & \checkmark & \checkmark & \textbf{0.8380} \\ 
    	\checkmark & \checkmark & \checkmark & \checkmark & & \checkmark & 0.8286 \\ 
    	\checkmark & & \checkmark & \checkmark & \checkmark & & 0.8286 \\ 
    	\checkmark & \checkmark & \checkmark & \checkmark & \checkmark & & 0.8263 \\
    	\checkmark & \checkmark & & \checkmark & \checkmark & \checkmark & 0.8263 \\
    	\checkmark & \checkmark & \checkmark & \checkmark & & & 0.8263 \\
    	\midrule
        \checkmark & & & & & & 0.7535 \\
    	& \checkmark & & & & & 0.7535 \\ 
    	 & & \checkmark & & & & 0.6901 \\ 
    	 & & & \checkmark & & & \textbf{0.7559} \\
    	 & & & & \checkmark & & 0.5798 \\ 
    	 & & & & & \checkmark & 0.6362 \\
    	\bottomrule
	\end{tabular} 
	\label{tab:eval_multidim}
\end{table}

We observe that the combination of all musical dimensions produces the best result for perceptual triplet prediction. This was expected, since having all dimensions activated provides the most informative representation. It can be observed that genre and era are very good indicatives for music similarity, since they are present in all the best combinations. This is further confirmed when evaluating each dimension separately. Somewhat surprisingly, the era dimension presents the higher rating, followed closely by the genre and mood dimensions. These results are a clear indicative that higher-level conceptualizations in music (genre, mood, and era) are individually more informative for music similarity, in comparison to mid/low-level notions (instrument, key, and tempo). This can be attributed to the fact that higher-level dimensions may already contain some information about them in their encoded representation. Still, the disentanglement of these low-level concepts provide a higher degree of informative cues about them, which benefits the entire combination.

\section{Conclusion}

In this paper, we introduce a multi-input deep architecture tailored for the disentanglement of both high-level semantic conceptualizations (genre, mood, and era), as well as mid/low-level descriptors (instrument, key, and tempo) of music from deep representations. We present a perceptual evaluation of music similarity using a disentanglement metric learning approach. The proposed architecture outperforms the baseline when evaluated with human-annotated similarity triplets from the Dim-sim dataset. Through the produced disentangled representations, we present a multi-dimensional analysis of perceptual music similarity, showing the influence of each musical dimensions individually. In this way, it is observed for the first time how a deep architecture models the multi-dimensionality of the human notion of music similarity.

\section*{Acknowledgements}
This research was partially supported by H2020 EU project AI4Media --  A European Excellence Centre for Media, Society and Democracy -- under Grand Agreement 951911.

\bibliographystyle{IEEEbib}
\bibliography{References.bib}

\end{document}